# The role of lattice thermal conductivity suppression by dopants from a holistic perspective

Shengnan Dai[1], Shijie Zhang[1], Ye Sheng[2], Erting Dong[3], Sheng Sun[1], Lili Xi[1], G. Jeffrey Snyder[4], Jinyang Xi[1,*], Jiong Yang[1,*]

1. Materials Genome Institute, Shanghai University, Shanghai 200444, China.
2. Department of Materials Science and Engineering, Southern University of Science and Technology, Shenzhen, 518055, China.
3. College of Materials Science and Engineering, Henan Institute of Technology, Xinxiang, Henan, 453000, China.
4. Materials Science and Engineering, Northwestern University, Evanston, Illinois 60208, USA.

**Abstract**

Dopants play an important role in improving electrical and thermal transport. In the traditional perspective, a dopant suppresses lattice thermal conductivity ($\kappa_L$) by adding point defect (PD) scattering term to the phonon relaxation time, which has been adopted for decades. In this study, we propose an innovative perspective to solve the $\kappa_L$ of defective systems — the holistic approach, i.e., treating dopant and matrix as a holism. This approach allows us to handle the influences from defects explicitly by the calculations of defective systems, about their changed phonon dispersion, phonon-phonon and electron-phonon interaction, etc, due to the existence of dopants. The $\kappa_L$ reduction between defective $M_xNb_{1-x}FeSb$ (M=V, Ti) and NbFeSb is used as an example for the holistic approach, and comparable results with experiments are obtained. It is notable that light elemental dopants also induced the avoided-crossing behavior. It can be further rationalized by a one-dimensional atomic chain model. The mass and force constant imbalance generally generates the avoided-crossing phonons, mathematically in a similar way as the coefficients in traditional PD scattering, but along a different direction in $\kappa_L$ reduction. Our work provides another perspective for understanding the mechanism of dopants influence in material's thermal transport.

## Introduction

In material science, the point defect (PD) is a type of imperfection in the crystal structure at the scale of individual atomic sites, and disrupts the locally regular arrangement of atoms. The structural, optical, electrical, and thermal properties of material can be tailored through the introduction of PD such as vacancy, interstitial, substitutional, and anti-site defect[1, 2, 3]. Defect engineering finds applications across various fields, such as thermoelectrics (TEs). One aspect of the PD enhancing the thermoelectric performance is the effective suppression in lattice thermal conductivity ($\kappa_L$) [4, 5, 6]. Historically, the PD model derived by Klemens[7, 8, 9, 10] is used to model the experimental trends of $\kappa_L$ versus point defect concentration. The traditional PD model is based on second-order perturbation theory, and adds a PD scattering mechanism to the phonon lifetime which represents the mass and strain fluctuation introduced by PD. The perturbation approach is schematically shown on the left side of Fig. 1, where the substitutional dopant is treated as independent individual separated from the matrix. The PD scattering rate ($\frac{1}{\tau_{PD}}$) and the phonon-phonon (PP) scattering rate of perfect structure ($\frac{1}{\tau_{undoped}^{PP}}$) are combined to derive the total scattering rates of defective structure ($\frac{1}{\tau_{tot}}$) via Matthiessen's rule[11]. Usually, the $\frac{1}{\tau_{PD}}$ is shown as the fourth power of frequency, i.e., $\frac{1}{\tau_{PD}} \sim A\omega^4$ [7]. The perturbational approach uses the phonon velocity of the defect-free system ($v_{undoped}$) under the assumption of unchanged phonon spectrum. Over decades, the PD model has been widely used as a cornerstone in solid thermal transport. It works well not only with the low concentration intrinsic defects and isotope as the cases when the PD model was originally proposed, but also with high concentrations and complete solid solutions[12, 13]. For example, Yang et al.[14] simulated the $\kappa_L$ reduction induced by mass and strain field fluctuation in $Zr_{0.5}Hf_{0.5}NiSn_{0.99}Sb_{0.01}$ compound using the PD model. Yu et al.[15] revealed that $(Nb_{0.6}Ta_{0.4})_{0.8}Ti_{0.2}FeSb$ has a larger $\kappa_L$ reduction than $Nb_{0.8}Ti_{0.2}FeSb$ due to the stronger PD scattering by fitting the PD parameter. Luo et al.[16] discovered that the effect of PD scattering on $\kappa_L$ reduction in alloyed triple half-Heusler is more apparent than other scattering mechanisms. In these cases, the solid solution induced-PD scattering is regarded as the main factor affecting the $\kappa_L$s of respective systems.

As mentioned, the PD model solely attributes defect-induced effect to the PD scattering mechanism, with an assumption that the PP interactions and the phonon dispersion is unchanged[17]. While this assumption holds for trace defect content, it becomes questionable in heavily doped or solid solution applications, as the high concentration defect inevitably disturb the properties of the matrix. On the other hand, the defects or solid solutions induced-effect in experiment is mostly determined through the fitted PD parameter[12, 13, 14]. However, there are many complex interfering factors in experiment, such as microstructure effects[18], grain boundaries[19] and dislocations[20] et al., which compromises the accuracy of analyzing the defect-induced effects. Besides, the additional effects such as the change in elastic properties due to defect strain[21] and electronic doping[22] are also not included in perturbational approach.

Based on the aforementioned considerations, we proposed an innovative perspective to solve the $\kappa_L$ of defective systems, which treats dopant and matrix as a holism and named as holistic approach. In other words, the traditional PD no longer exists; instead, it was viewed as a part of the matrix. In the $\kappa_L$ treatment, only the intrinsic and treatable properties, such as the lattice vibration and the PP and electron-phonon (EP) scattering, are considered in the holistic approach. This approach makes it possible to deal with the mechanisms induced by dopants *in-silico* without being disturbed by other effects in experiments[23, 24]. The schematic plot of the holistic approach is displayed on the right of Fig. 1. Unlike the traditional perturbational approach, the holistic approach incorporates the dopant-induced alterations in 2$^{nd}$ and 3$^{rd}$ IFCs. These alterations mainly reflected in the phonon dispersion (e.g., $v_{doped}$) and the scattering rate ($\frac{1}{\tau_{doped}^{PP}}$). In addition, the electron-phonon scattering rate ($\frac{1}{\tau_{doped}^{EP}}$) caused by the interaction of the introduced charge with phonon is also considered. The individual PD scattering rate ($\tau_{PD}$), is not considered anymore. In practice, the number of atoms considered in the simulation is limited by computational power; the good thing is that current thermoelectric material systems with heavy-doped content are in the capability of the density functional theory (DFT)[17]. On the basis of allowable computing power, the holistic approach even enables to mimic the intrinsic property of system with more complex defect or infinite atoms. Consequently, this study employs the holistic approach to investigate the $\kappa_L$ of defective systems.

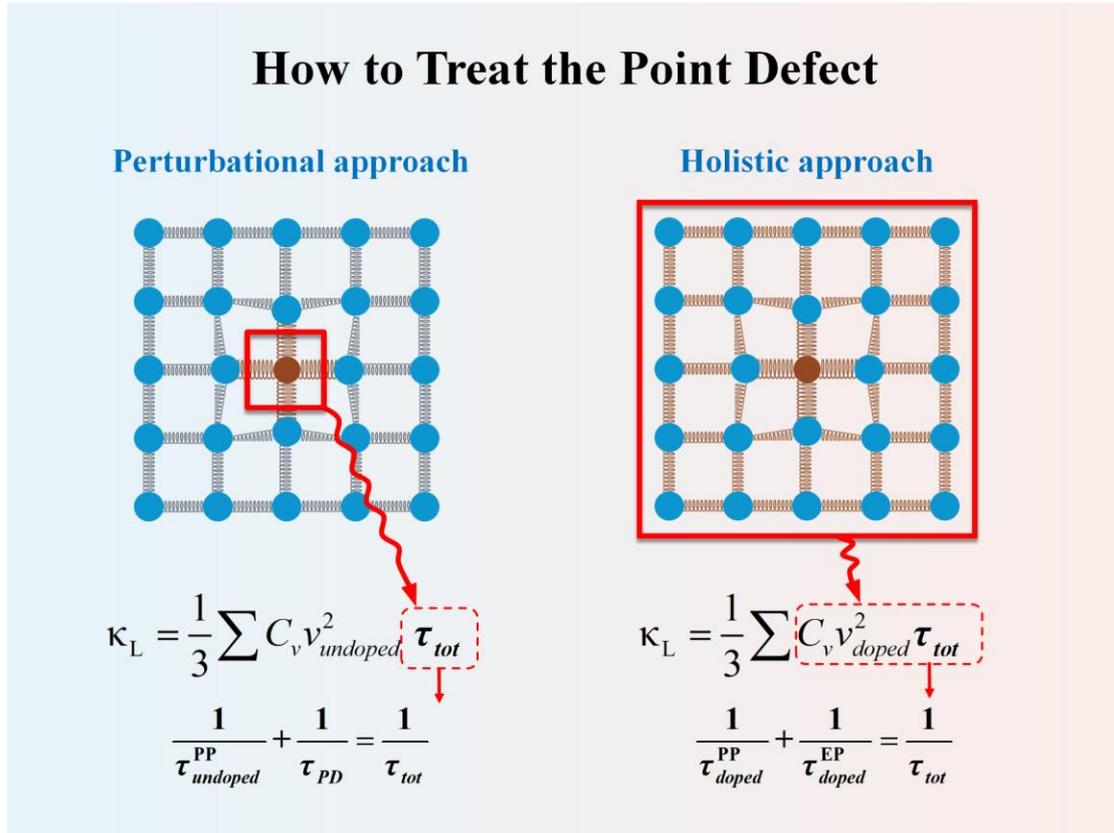

**Fig. 1.** The two approaches to treat point defect on $\kappa_L$ in crystal, the traditional perturbational approach on the left and the holistic approach proposed in this work on the right. The scattering rates from different phonon scattering mechanisms considered by each of approaches are shown at the bottom of the schematic, respectively. Where $C_v$ is the heat capacity, $v$ is the phonon group velocity and $\tau$ is the phonon relaxation time.

The half-Heusler (HH) compounds have emerged as the potential TE materials since the year of 2000[25] due to the good TE properties. The HH compound has a cubic MgAgAs-type structure with the space group $F\bar{4}3m$, and they are typically heavy doped in applications through substitution[26, 27], which exhibit a marked reduction in $\kappa_L$. In this work, NbFeSb was selected as the host compound, and the Ti substituting on Nb site is the main dopant in interest. The $\kappa_L$ reduction of the Ti-doped NbFeSb is abnormally large, especially considering the light atomic weight of the Ti atom. Furthermore, V substituting on the Nb site is also considered in order to compare the different of aliovalent and isoelectronic dopants. For all the systems, we adopt the holistic approach to consider the changes in 2$^{nd}$ and 3$^{rd}$ IFCs brought by dopants, as well as the effects of EP interaction. The calculation results show that even with light element-doping, the $\kappa_L$ reduction can reach nearly 40% at 300K when 25% doped. The aliovalent dopant Ti can further reduce the $\kappa_L$ to 60% due to the EP interaction. All the results are obtained without considering the PD scattering. Nevertheless, the $\kappa_L$ reduction obtained by the holistic approach, which only considering the changed phonon dispersion, PP and EP interaction, essentially covered the experimental $\kappa_L$ reduction. Furthermore, a one-dimensional four-atom (1d4A) chain model is proposed, and the numeral results demonstrate that the mass and force constant imbalance, which is attributed to PD scattering rate in the perturbational approach, induce the avoided-crossing behavior in phonon dispersion. The quantitative analysis indicated that the degree of avoided-crossing induced by mass and force constant fluctuations is similar to the numerical representation of the traditional PD scattering rate, but the two stemmed from different approaches. This work provides a comprehensive understanding of the role of dopants on $\kappa_L$ reduction and presents a new perspective for dealing with the thermal transport of defective system.

**Results**

**Thermal transport properties**

Based on the holistic approach, we calculated the $\kappa_L$ of NbFeSb with aliovalent Ti-doping and isoelectronic V-doping under a series of concentrations, i.e., $M_xNb_{1-x}FeSb$ (M=Ti or V, x= 0.037, 0.125, 0.25). For aliovalent doped $Ti_xNb_{1-x}FeSb$, the $\kappa_L$s considering only PP interaction ($\kappa_{PP}$) and both PP and EP interaction ($\kappa_{PP+EP}$) are calculated. For isoelectronic doped $V_xNb_{1-x}FeSb$, the $\kappa_L$s consider only the PP interaction. The cells with dopants containing 12, 24, and 81 atoms are constructed to accommodate the 25%, 12.5% and 3.7% doped contents, respectively. The finite displacement method is used to calculate the 2$^{nd}$ and 3$^{rd}$ IFCs, with the corresponding supercells containing 324 (3×3×3×12 atoms unit-cell), 192 (2×2×2×24 atoms unit-cell), and 648 (2×2×2×81 atoms unit-cell) atoms, more details can be found in the Methods section.

In principle, the calculations of the EP interactions on $\kappa_L$ require the EP coupling matrix. The computational effort in calculating the EP coupling matrix element increases exponentially as the number of atoms rises. We thus developed a theoretical approach under the acoustic phonon long-wavelength limit[28, 29, 30], which replaces the EP coupling matrix element with deformation potential and bulk modulus. We referred to this algorithm as Thermal Transport from the Electron-Phonon interaction (TTEP). Below is the formula for simplified electron-phonon scattering rate, the derivation details are in SI:

$$\frac{1}{\tau_{\mathbf{q}\lambda}^{EP}} = \frac{\pi D^2}{VB} \sum_{mn,\mathbf{k}} \omega_{\mathbf{q}\lambda} \left( f_{m\mathbf{k}+\mathbf{q}} - f_{n\mathbf{k}} \right) \delta \left( \varepsilon_{n\mathbf{k}} - \varepsilon_{m\mathbf{k}+\mathbf{q}} - \hbar\omega_{\mathbf{q}\lambda} \right), \qquad (1)$$

where $D$ is the deformation potential, $B$ is the bulk modulus and $V$ is the volume of the unit-cell. The TTEP algorithm is adopted for the aliovalent Ti-doped NbFeSb. The deformation potential and bulk modulus of each doped systems are calculated in the same way as for Jin et al.[31], and the values are displayed in Table S1. There is little difference in values among the doped systems, because they share the same matrix. Accurate EP interaction is calculated on two systems, x(Ti) = 0.125 and x(Ti) = 0.25. As illustrated in Fig. 2a, the $\kappa_{PP+EP}$ of each system obtained by the efficient TTEP algorithm is in general agreement with that obtained by the accurate EP algorithm. Therefore, it is reasonable to believe that the Ti$_{0.037}$Nb$_{0.967}$FeSb's $\kappa_{PP+EP}$ (81-atom unit cell) obtained by TTEP algorithm is reliable, which is challenging to achieve by accurate EP algorithm.

To demonstrate the effect of the dopant on the thermal transport properties of the matrix, we define the $\kappa_L$ reduction rate between doped Ti(V)$_x$Nb$_{1-x}$FeSb (x=0.037, 0.125,

0.25) and the matrix NbFeSb as η: $\eta = \frac{\kappa_{doped} - \kappa_{undoped}}{\kappa_{undoped}}$. The η for Ti(V)$_x$Nb$_{1-x}$FeSb versus dopant content at 300 K is showed in Fig. 2a. With an increase in Ti content, the $\kappa_{PP}$ keep decreasing. After considering the EP interaction, the change ratio of $\kappa_L$ is further reduced. The $\kappa_{PP}$ ($\kappa_{PP+EP}$) of 12.5% Ti-doped has a reduction of 31% (44%) compared to NbFeSb. When the Ti-doping concentration reaches 25%, the reduction in $\kappa_{PP}$ can reach 44% and the reduction in $\kappa_{PP+EP}$ can reach 63%. $\kappa_{PP}$ is only reduced by 26% and $\kappa_{PP+EP}$ is only further reduced to 34% as a result of the 3.7% Ti-doping. The $\kappa_L$ reduction after considering EP interaction approaches the experimental value of Fu et al. [32], proposing that first-principles calculations can offer insights into the role of dopants on thermal transport. Meanwhile, Villoro et al.[23] demonstrated that Ti doped NbFeSb samples emerge unusual grain boundary and complex phase, which also have effects on the thermal conductivity. The variation of $\kappa_L$ reduction versus content of V-doping is slower than that of Ti-doping due to the lack of EP coupling. The holistic approach's $\kappa_L$ vs dopant content curve exhibits a *U*-shaped trend and is comparable to the traditional perturbational approach[33], but under totally different methodologies. In order to reveal the role of $\kappa_L$ suppression by dopants, we selected the Ti(V)$_{0.25}$Nb$_{0.75}$FeSb which has the largest doping content in this paper for analysis (pink area). Figure 2b shows the calculated phonon frequency-dependent $\kappa_L$ for the two doped systems and NbFeSb at 300 K. After the introduction of dopant, the $\kappa_L$ reduction is mainly at 0-4 THz when only considering the PP interaction. The $\kappa_L$ reduction from EP interaction is mainly at 0-3 THz, corresponding to acoustic phonon region. To further distinguish how the isoelectronic and aliovalent dopants dominate these two factors, we recalculated the $\kappa_L$ using $\kappa_L = \frac{1}{3}\sum C_v v^2 \tau$ by considering both changes in $v$ and $\tau$, only changing $\tau$ (with $v$ from the matrix NbFeSb) and only changing $v$. Figure 2c illustrates that changing both $v$ and $\tau$ contributes to the $\kappa_L$ suppression. Hence, the role of isoelectronic and aliovalent dopants on the two factors are displayed in the following.

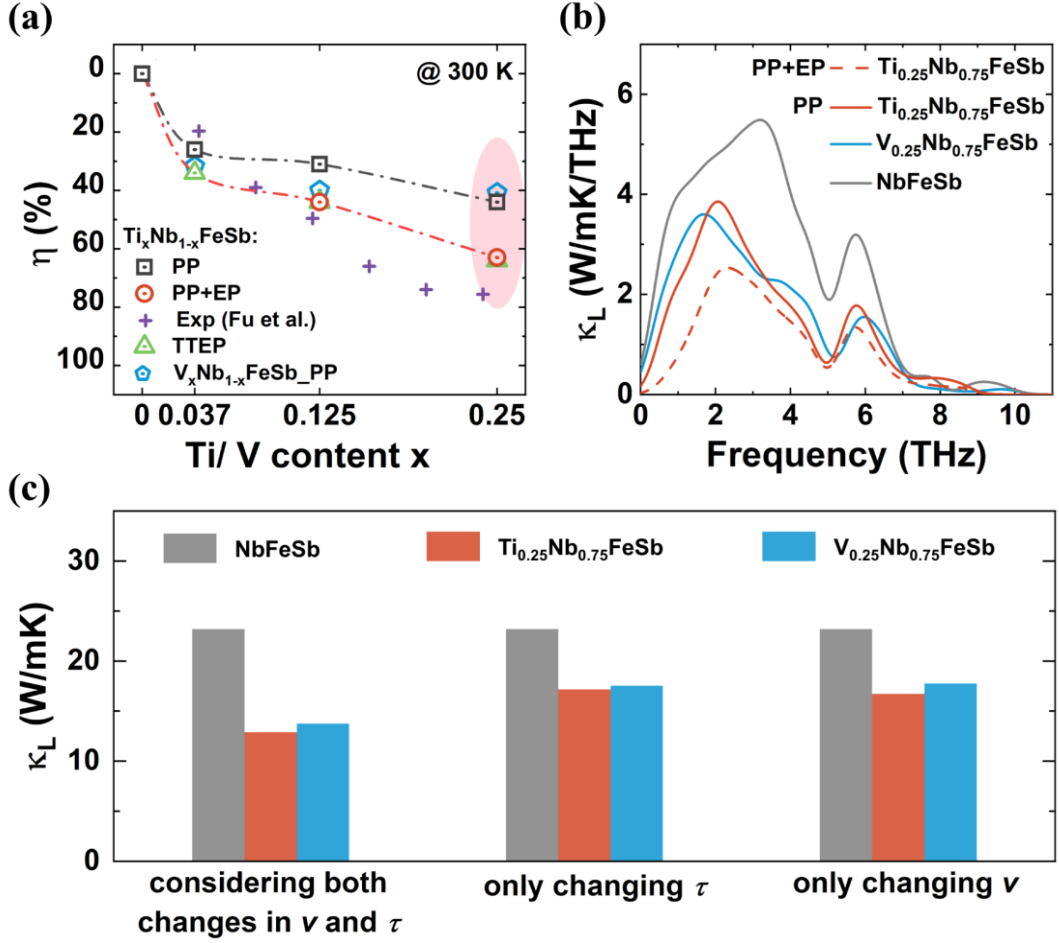

**Fig. 2.** (a) The $\kappa_L$ reduction rate (η) versus nominal Ti/V content x for Ti(V)$_x$Nb$_{1-x}$FeSb at 300K. The black squares, red circles and green triangles are the values of Ti$_x$Nb$_{1-x}$FeSb under PP interaction only, EP interaction and TTEP algorithm, respectively. And the purple crosses are the experimental values of Ti$_x$Nb$_{1-x}$FeSb by Fu et al. The blue pentagons are the values of V$_x$Nb$_{1-x}$FeSb under PP interaction. (b) Phonon frequency-dependent $\kappa_L$ for Ti(V)$_{0.25}$Nb$_{0.75}$FeSb and NbFeSb at 300K. Where the solid line represents $\kappa_{PP}$, and the dotted line represents $\kappa_{PP+EP}$. (c) The calculated $\kappa_L$ for Ti(V)$_{0.25}$Nb$_{0.75}$FeSb and NbFeSb at 300K by considering both changes in $v$ and $\tau$, only changing $\tau$ and only changing $v$.

**Dopant-induced changes in the phonon group velocity**

The phonon group velocity factor is covered first. Since the $\kappa_L$ reduction mainly happens in the range of 0-4 THz (shown in Fig. 2a), the frequency dependence of phonon group velocity for Ti(V)$_{0.25}$Nb$_{0.75}$FeSb compared to NbFeSb in this range are presented in Fig. 3a. Figure S1 depicts the phonon group velocity results for all the V-/Ti-doped systems. The phonon dispersions of all systems are shown in Fig. S2. In order to observe the effect of dopants on the phonon spectrum more intuitively, the unfolding phonon spectrum[34] of cell-expansion systems are displayed in Fig. S3. Compared to only defect effect in isoelectronic V-doping, both electron and defect have an impact on the phonon dispersion for aliovalent Ti-doping. At 0-2 THz (the red region), the $v$ of $Ti_{0.25}Nb_{0.75}FeSb$ is lower than $V_{0.25}Nb_{0.75}FeSb$ and NbFeSb (Fig. 3a). In this region, the difference of phonon group velocity means the softening of the acoustic phonon branches around the Brillouin zone center, the transverse acoustic branches in this case. Figure 3b demonstrates that there is a distinct decrease of TA modes in $Ti_{0.25}Nb_{0.75}FeSb$ (the red region), while no change is observed for the V-doped. We further calculate the low-frequency phonon dispersions for aliovalent doped $Zr(Hf)_{0.25}Nb_{0.75}FeSb$ and isoelectronic doped $Ta_{0.25}Nb_{0.75}FeSb$ (Fig. S4). Similar with the cases in the Ti-doped, the softening of the TAs are observed in $Zr(Hf)_{0.25}Nb_{0.75}FeSb$, not in the Ta-doped. Interestingly, an artificial calculation for NbFeSb with one less electron (same electron count with $Ti_{0.25}Nb_{0.75}FeSb$) is also carried out, as also shown in Fig. S4. The TAs of one-electron-less NbFeSb sit on top on the curves for $Ti_{0.25}Nb_{0.75}FeSb$, confirming that the softening of the TA modes around the zone center in NbFeSb is due to the aliovalent doping.

In the phonon region 2-3 THz (the yellow region), both the doped systems exhibit another suppression in $v$ in comparison to NbFeSb. As shown in Fig. 3b, the acoustic branches of NbFeSb cross normally with optical branches, whereas there is a strong avoided-crossing behaviors happening around 2.5 THz in $Ti_{0.25}Nb_{0.75}FeSb$ and $V_{0.25}Nb_{0.75}FeSb$, which impeding the phonon propagation in this region. The avoided-crossing behavior was originally reported in the structure with heavy element dopant[17, 35, 36], a similar phenomenon was also discovered in Hf- and Ta-doping (given in Fig. S5). Besides, the doping with Zr which is in similar mass to Nb (Fig. S5) also show weak avoided crossing. Such behavior also reduces the phonon group velocity of acoustic phonons (~ 3 THz). This is further evidence that dopant affects the phonon dispersion[17], as opposed to the traditional point defect model that the defect is an independent entity and the phonon structure unchanged after doping. Furthermore, in contrast to the V-doping system's dispersed optical branch, Ti$_{0.25}$Nb$_{0.75}$FeSb's has a narrower frequency range, which predicts a larger scattering phase space.

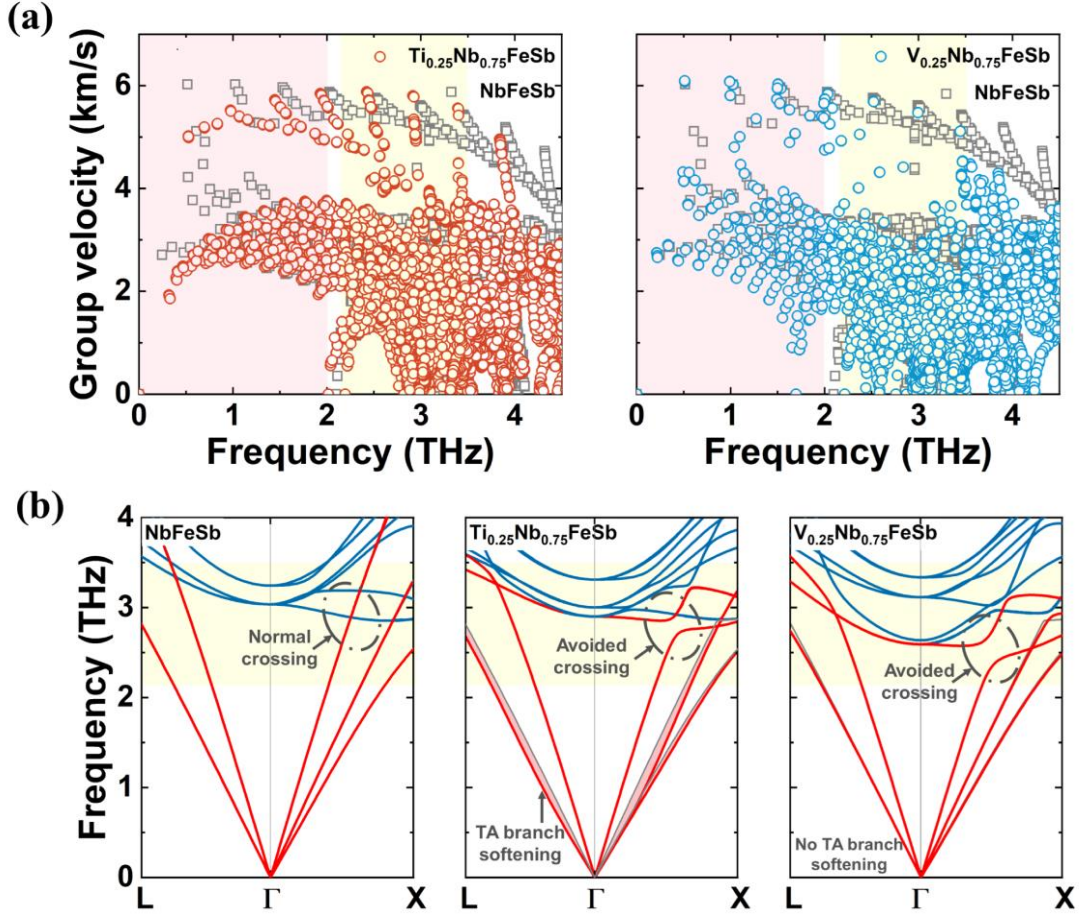

**Fig. 3.** (a) The phonon group velocity as the function of phonon frequency for Ti$_{0.25}$Nb$_{0.75}$FeSb and V$_{0.25}$Nb$_{0.75}$FeSb compared to NbFeSb (0-4.5 THz). (c) The phonon dispersion for NbFeSb, Ti$_{0.25}$Nb$_{0.75}$FeSb and V$_{0.25}$Nb$_{0.75}$FeSb (0-4 THz). The avoided-crossing behavior comparison was circled. The TA branch softening behavior is compared with the NbFeSb TA branches (gray line).

It is interesting that the light element dopant induces the anomalous avoided-crossing behavior between the acoustic and optical branches, which violate the common intuition that only heavy dopants will introduce this behavior. In order to rationalize the result, we built a 1d4A chain model to simulating the lattice vibration of the doped system. Figure 4a illustrates the crystal structure of the doped M$_{0.25}$Nb$_{0.75}$FeSb (M=Nb,

Ti, V, Zr, Hf, Ta). The blue ball represents the dopant replacing the Nb site. In the simplified 1d4A model (as shown in Fig. 4b), each host Nb atom has mass $M_1$ and the Nb interact with dopant through the force constant $K_1$. The interstation Fe atom has mass $M_2(M_4)$ and interact with the host Nb atom through force constant $K_2$, interact with the dopant through force constant $K_3$. The negative integrated projected crystal orbital Hamilton population (–IpCOHP) [37, 38, 39] was selected to represent the interaction K in this study. All the parameters adopted in the 1d4A model can be found in Table S2. It is worth noting that, the next-nearest-neighboring interaction, $K_1$ in our case, is much weaker than the nearest-neighboring ones $K_2$ and $K_3$. The fact that nearest-neighboring force constant is much stronger is more general than the previous model discussing the case in the caged compounds[35], where $K_1 \gg K_2$. The derivation detail of the dispersion solution for 1d4A chain model is shown in the SI. The dispersion solutions for $M_{0.25}Nb_{0.75}FeSb$ (M=Nb, Ti, V) using the 1d4A chain model are illustrated in Fig. 4c, and Zr, Hf, Ta doping is given in Fig. S6. Employing $\omega_q/\omega_{std}$ ($\omega_{std} = \sqrt{\frac{K_2}{M_1}}$, where $K_2$ and $M_1$ belong to the host NbFeSb, as shown in Fig. 4b), the dimensionless dispersion solution is built. In the undoped case of NbFeSb ($M_1=M_3$, $K_2=K_3$), the dispersion solution of the 1d4A chain (Fig. 4c) reproduces the normal crossing behavior that exist in the DFT phonon dispersion (Fig. 3b). However, the dispersion solution of 1d4a chain for $Ti_{0.25}Nb_{0.75}FeSb$ ($M_1>M_3$, $K_2>K_3$) arose a gap, which is consistent with the avoided-crossing behavior in DFT calculation. So did the $V_{0.25}Nb_{0.75}FeSb$. The aforementioned indicates that light element dopants can also induce avoided-crossing behavior, which impede phonon propagation and limit phonon velocity. Naturally, the Hf and Ta doped systems' dispersion solution ($M_1<M_3$, $K_2>K_3$) also has the avoided-crossing as traditionally believed, showing in Fig. S8. The $Zr_{0.25}Nb_{0.75}FeSb$'s dispersion solution ($M_1\approx M_3$, $K_2>K_3$) also keeps the avoided crossing with a smaller gap, consistent with the Fig. S5. In summary, the 1d4A chain model developed in this work reproduces the avoided-crossing behavior more intuitively. It also suggests that the phenomenon is originated form an imbalance of force constants and masses, not only due to the heavy mass doping. A more versatile 1d4A model will be discussed in the **Discussion** section.

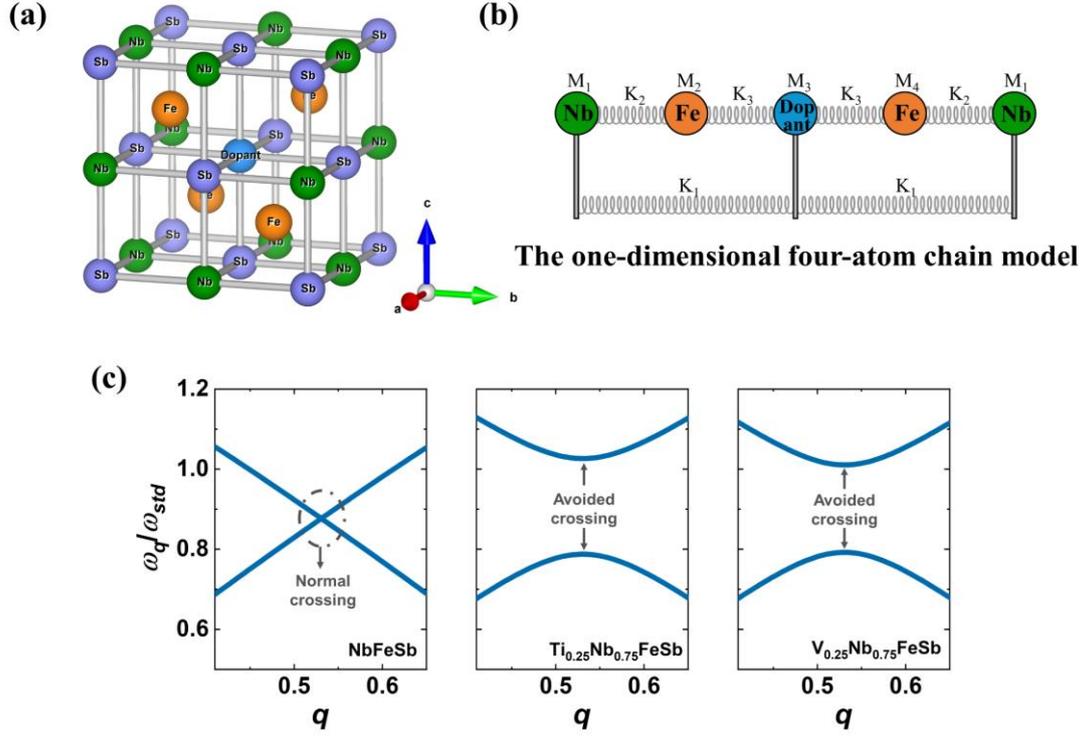

**Fig. 4.** (a) The crystal structure of the doped $M_{0.25}Nb_{0.75}FeSb$ (M=Nb, Ti, V). The blue ball represents the dopant. (b) A simplified 1d4A model describing the interaction between the host Nb atom, the interstation Fe atom and the dopant. The force constant between Nb and dopant is $K_1$, Nb and Fe is $K_2$, Fe and dopant is $K_3$. (c) The dimensionless dispersion solutions for $M_{0.25}Nb_{0.75}FeSb$ (M=Nb, Ti, V) using the simplified 1d4A model in (b).

**Dopant-induced changes in the scattering rate**

As shown in Fig. 2b, the variations in scattering rate are also important to the $\kappa_L$ reduction. Only the intrinsic PP scattering rate ($\frac{1}{\tau_{q\lambda}^{PP}}$) and the EP scattering rate ($\frac{1}{\tau_{q\lambda}^{EP}}$) stemming from aliovalent doping are considered in this work. These two together constitute the total scattering rate ($\frac{1}{\tau_{q\lambda}}$) through the Matthiessen's rule, delineated as $\frac{1}{\tau_{q\lambda}} = \frac{1}{\tau_{q\lambda}^{PP}} + \frac{1}{\tau_{q\lambda}^{EP}} + \cdots$. The two figures in Fig. 5a above shown the PP scattering rate of

Ti$_{0.25}$Nb$_{0.75}$FeSb and V$_{0.25}$Nb$_{0.75}$FeSb compared to NbFeSb, and the Fig. S9 displayed the PP scattering rate vary with frequency for other systems. Both aliovalent Ti$_{0.25}$Nb$_{0.75}$FeSb and isoelectronic V$_{0.25}$Nb$_{0.75}$FeSb illustrate slightly higher PP scattering rate than NbFeSb. In order to quantitatively measure the difference of three systems' scattering rates, the average scattering rates were calculated by using the square of phonon group velocity as weighting[40]. The average PP scattering rate of all systems are displayed in Table S3, and the average PP scattering rate of x=0.25 is 0.08 THz (NbFeSb), 0.104 THz (Ti-doped) and 0.1 THz (V-doped). The role of aliovalent and isoelectronic dopants on PP scattering rate are similar. The scattering rate in PP interaction processes can be expressed as:

$$\tau_{PP}^{-1} = \frac{\hbar\pi}{8}\sum_{\lambda'\lambda''}\left|\Phi_{\lambda\lambda'\lambda''}\right|^2\left[\left(n_{\lambda'}+n_{\lambda''}+1\right)\delta\left(\omega_\lambda-\omega_{\lambda'}-\omega_{\lambda''}\right)+2\left(n_{\lambda'}-n_{\lambda''}\right)\delta\left(\omega_\lambda-\omega_{\lambda'}+\omega_{\lambda''}\right)\right]$$

$\Phi_{\lambda\lambda'\lambda''}$ is the relevant quantity for 3$_{nd}$ IFCs that include the anharmonic interaction. The delta function is energy conservation and the expression enclosed in bracket related to phonon scattering phase space. Thus, the PP scattering rate can be divided into the contribution of the scattering phase space, and the anharmonic effect, usually reflected by the Grüneisen parameter. Figure 5b shows the scattering weighted phase space (W)[41] of Ti$_{0.25}$Nb$_{0.75}$FeSb and V$_{0.25}$Nb$_{0.75}$FeSb compared with NbFeSb, respectively. The W of X$_{0.125}$Nb$_{0.875}$FeSb is depicted in Fig. S10. The W of V$_{0.25}$Nb$_{0.75}$FeSb is only slightly higher than that of NbFeSb in the middle frequency region. The difference is that the W of Ti$_{0.25}$Nb$_{0.75}$FeSb is significantly higher than that of NbFeSb, especially at 0-7 THz. This also confirms the previous conjecture about scattering phase space at the phonon dispersion. The aliovalent dopant induces an increase in the scattering phase space of the matrix. This demonstrates once again the *Han et al.*[17] that the aliovalence dopant softs the high-energy optical phonons, and further expands the scattering phase space. In addition, the comparison of the Grüneisen parameter of both system with NbFeSb is given in Fig. 5c, and other systems' Grüneisen parameters are shown in Fig. S11. The Grüneisen parameter 's absolute value reflects the magnitude of the anharmonic effect. While the Grüneisen parameters of isoelectronic-doped V$_{0.25}$Nb$_{0.75}$FeSb are slightly larger than undoped NbFeSb, the aliovalent Ti dopant decreases the Grüneisen parameters.

Next, as illustrated at the bottom of Fig. 5a, $Ti_{0.25}Nb_{0.75}FeSb$ has sufficiently large EP scattering rates which are even comparable to the PP scattering rates at the low-frequency range, indicating a strong EP interaction. The EP interaction causes $Ti_{0.25}Nb_{0.75}FeSb$'s $\kappa_L$ to reduce from 44% to 63%. The PP and EP scattering rates of $Ti_{0.125}Nb_{0.875}FeSb$ shown in Fig. S12, while the average scattering rates are listed in Table S3. As shown in Table S3, the EP scattering rate declines with the content of aliovalent dopant. This phenomenon arises from the gradual penetration of Fermi level in the valence band with the dopant content, and there is an augmentation in the electronic density of states at the Fermi level, improving the EP interaction[40]. Furthermore, we have investigated the EP scattering obtained through the TTEP algorithm, as depicted in Fig. S13. The average scattering rate of TTEP is 0.048 THz, which is close to the value from the accurate EP scattering rate (0.052 THz). The close values of the average EP scattering rate obtained by both algorithms rationalize that the Ti-doped $\kappa_{PP+EP}$ predicted by TTEP as shown in Fig. 2a is reasonable. The very low frequency range is where the majority of the difference between the EP scattering rates determined by the TTEP algorithm and the accurate algorithm. The difference stems from the fact that the TTEP uses the deformation potential approximation, which is theoretically comparable to the effect of LA phonon. But the TA phonon usually has a larger EP coupling matrix element[40] and higher density of states than LA phonon. The difference in the low frequency range, however, has little effect on $\kappa_L$ reduction.

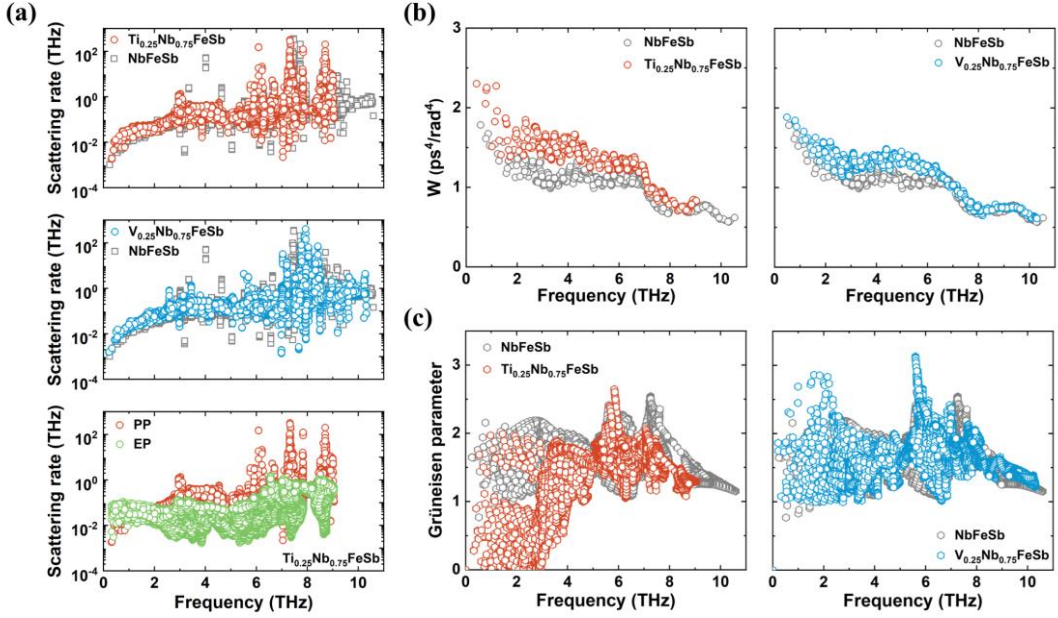

**Fig. 5.** (a) The PP scattering rate of $Ti_{0.25}Nb_{0.75}FeSb$ and $V_{0.25}Nb_{0.75}FeSb$ as the function of phonon frequency compared to NbFeSb, respectively. The third figure is the phonon scattering rate from PP interactions and EP interactions as the function of phonon frequency for $Ti_{0.25}Nb_{0.75}FeSb$. (b) The phonon frequency dependence of PP scattering weighted phase space for $Ti_{0.25}Nb_{0.75}FeSb$ and $V_{0.25}Nb_{0.75}FeSb$. (c) The phonon frequency dependence of Grüneisen parameter for $Ti_{0.25}Nb_{0.75}FeSb$ and $V_{0.25}Nb_{0.75}FeSb$.

The above discussed are the results of the temperature at 300K. The temperature-dependent $\kappa_L$s for $M_xNb_{1-x}FeSb$ are displayed in Fig. S14. The $\kappa_L$ considering only PP interaction as a function of temperature obey the typical $T^{-1}$ law. After considering the EP interaction, the curves of $Ti_{0.25}Nb_{0.75}FeSb$ and $Ti_{0.125}Nb_{0.875}FeSb$'s $\kappa_L$ with temperature deviate from $T^{-1}$ law to $T^{-0.8}$ and $T^{-0.9}$, respectively. This is due to the different temperature dependence in the EP and PP scattering rates. The phonon Bose-Einstein distribution in PP scattering rate is proportional to temperature ($n \propto T^1$) under the high-temperature limit, but the Fermi-Dirac distribution in EP scattering rate is almost temperature-independent[40]. The more content of aliovalent dopants, the stronger the EP interaction and the more significant the deviation.

**Discussion**

The PD model under the perturbation theory was proposed in the historical context of trace doping on intrinsic semiconductors or insulators, such as KCl doped with KI ($5\times10^{19}$ cm$^{-3}$)[9]. Since the interatomic forces were not available, the physical quantities associated with PD were assumed to be the perturbation to the host, which producing mass and strain-field difference. This is the standard practice for dealing with unknown systems in quantum mechanics. The PD model based on the perturbation theory, which has been widely used up to now. Numerous studies have also demonstrated that it provides a good qualitative explanation for variations in $\kappa_L$ caused by PD[12, 13, 16]. But just as any physical approach is used to ensure the correctness of its theoretical premise, the perturbational approach, which only considers the PD scattering rate, is broken in the case of heavily doped HH[17]. The holistic approach as well as the numerical treatment in this study, offers a fresh viewpoint for handling of heavily doped HH and even heavily doped thermoelectrics. Similar treatment, i.e., the consideration of variations on phonon, phonon-phonon interactions by dopants, has also been employed in our group's past work, like doped Mg$_{2.75}$X$_{0.25}$Sb$_2$ [X = Ca; Yb] [42] and Bi$_2$Te$_3$-based compounds[43].

With accurate simulation results based on the holistic approach, we can get a clear picture of how the dopant specifically affects the thermal conductivity of the host, such as the phonon group velocity, scattering phase space, anharmonic and electron-phonon interaction, further enhancing the understanding of defect-induced $\kappa_L$ reduction. For example, *Han et al.*[17] study combined the first-principles calculation using holistic approach and the INS technique to confirm that dopants modulate the phonon dispersion. The 1d4A model with the parameters from the doped NbFeSb demonstrates that the imbalance of the mass and force constant induced by the dopants will introduce the avoided crossing behavior. To make the model more general, we remove the next-nearest neighboring interaction (K$_1$), and introduce only one dopant atom (M$_2$) out of the four host atoms (M$_1$), i.e., mimicking the 25% doping content (Fig. 6a). The relevant derivation can be seen in SI. The versatile model is not only adapted to the HHs, but also any doped systems in general. As discussed above, the avoided crossing does form as long as there exists imbalance of the mass and force constant fluctuation. In order to quantitatively reveal the relation between vibration frequency difference and the mass/force constant fluctuation, we define the $\Delta M = \frac{3}{4}\left(\frac{M_1-\overline{M}}{\overline{M}}\right)^2 + \frac{1}{4}\left(\frac{M_2-\overline{M}}{\overline{M}}\right)^2$, $\Delta K = \frac{3}{4}\left(\frac{K_1-\overline{K}}{\overline{K}}\right)^2 + \frac{1}{4}\left(\frac{K_2-\overline{K}}{\overline{K}}\right)^2$, $\overline{M}$ and $\overline{K}$ are the average mass and force constant. This form refers the mass and strain variation in the PD model. Moreover, the illustration of the vibration frequency difference ($\Delta\omega$) is displayed in the right corner of Fig. 6b, where $\omega_1^{max}$ is the maximum of the first dispersion solution and $\omega_2^{min}$ is the minimum of the second dispersion solution, and $\Delta\omega = \omega_2^{min} - \omega_1^{max}$. By setting the $M_1$, $K_2$ of the host as constant and the $M_2$, $K_3$ of the dopant as variable value, the square of frequency difference ($\Delta\omega^2$) varies with $\Delta M$ and $\Delta K$ is illustrated in Fig. 6b. The result reveals a good linear relationship between $\Delta\omega^2$ and $\Delta M$ ($\Delta K$), respectively. The force constant fluctuation has a stronger effect than mass fluctuation. The degree of avoided-crossing varies with mass and force constant fluctuations is similar to the description of PD scattering rate in PD model[13], but their physical grounds are different. The $\Delta M$ and $\Delta K$ determine the intensity of PD scattering in the PD model. In our versatile 1d4A chain model, however, they determine the degree of avoided-crossing, which control the thermal transport by the impact on the phonon dispersion characteristics, such as phonon velocity. In this sense, the fact that the $\kappa_L$ suppression in the defective system is directly caused by the influence of phonon dispersion by the dopants.

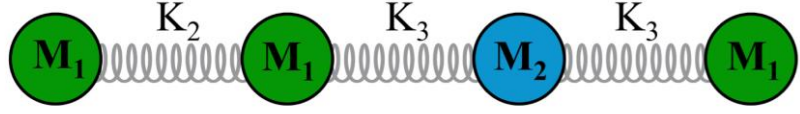

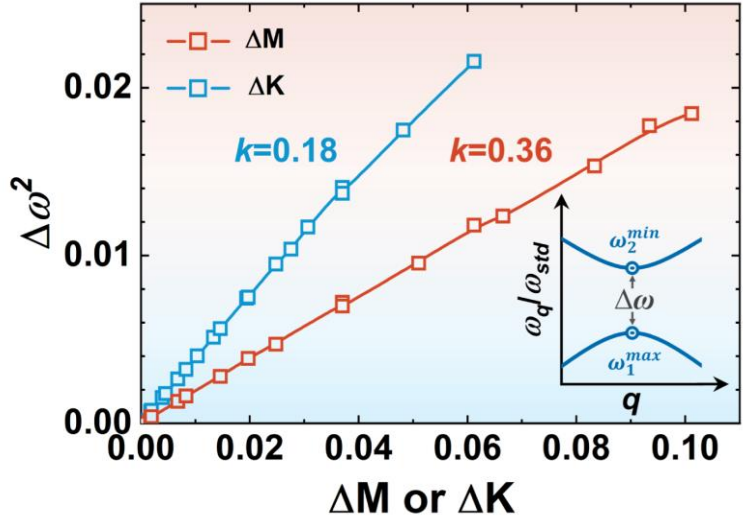

**Fig. 6.** (a) The versatile one-dimensional four-atom chain model that degenerates from the previously established model. (b) The dependence curves of mass (ΔM) and force constants (ΔK) fluctuations on square of frequency difference (Δω) determined by using the versatile 1d4A model. The schematic of Δω is shown in the right corner.

In summary, in contrast to the traditional perturbational approach, this work proposed the holistic approach which treats the PD and the surrounding environment as a holism. Light element (Ti, V) doped NbFeSb was chosen as a study case. We discover that light elements can also induced the avoided-crossing behavior in phonon diagram, which soften the acoustic phonon and reduce the $\kappa_L$. For isoelectronic dopants, the phonon dispersion of V-doped NbFeSb produces the avoided-crossing behavior and the phonon group velocity decreases, the scattering rate also decreases slightly compared to NbFeSb's. For aliovalent dopants, Ti-doped NbFeSb's scattering phase space is significantly elevated, but the scattering rates do not differ much from that of

NbFeSb due to the reduction of the Grüneisen constant, its phonon group velocity is reduced by the softening of TA branch and avoided-crossing behavior. Furthermore, according to the lattice dynamic model provided in this work, it is demonstrated that fluctuations in mass and force constant brought by the dopant affect the lattice vibrations by generating avoided-crossing. The quantitative result suggests that these fluctuations have a linear relation with the degree of avoided crossing, in the similar form as in the intensity of point defect scattering in the traditional PD model. Our work not only visualizes the effect of dopants on the intrinsic physical quantities of thermal transport more intuitively through the holistic approach, but also provides an innovative strategy to dealing with thermal transport in defective systems. The versatile 1d4A model can be extended to a 1djA (one-dimension j-atom) model to further discuss the association between PD content and frequency difference.

**Methods**

**The scattering rate under different scattering mechanisms**

In the framework of Boltzmann transport equation (BTE), the lattice thermal conductivity tensor can be expressed as[44]:

$$\kappa_L^{\alpha\beta} = \frac{1}{k_B T^2 NV} \sum_{\mathbf{q}\lambda} \left(\hbar\omega_{\mathbf{q}\lambda}\right)^2 \left(1+n_{\mathbf{q}\lambda}\right) v_{\mathbf{q}\lambda}^\alpha v_{\mathbf{q}\lambda}^\beta \tau_{\mathbf{q}\lambda}, \qquad (2)$$

where $\alpha$ and $\beta$ are the Cartesian directions, $k_B$ is the Boltzmann constant, $N$ is the number of **q** points in the first Brillouin zone, and $V$ is the unit cell volume. $\omega_{\mathbf{q}\lambda}$, $n_{\mathbf{q}\lambda}$, $v_{\mathbf{q}\lambda}$, and $\tau_{\mathbf{q}\lambda}$ are the frequency, Bose-Einstein occupancy, group velocity and relaxation time corresponding to a phonon mode with wave vector **q** and branch $\lambda$, respectively. The total phonon scattering rates consider the contribution from different scattering mechanisms according to Matthiessen's rule[11]

$$\frac{1}{\tau_{\mathbf{q}\lambda}} = \frac{1}{\tau_{\mathbf{q}\lambda}^{PP}} + \frac{1}{\tau_{\mathbf{q}\lambda}^{EP}} + \cdots, \quad (3)$$

This study focuses on the phonon-phonon (PP) scattering rates ($\frac{1}{\tau_{\mathbf{q}\lambda}^{PP}}$) and the EP scattering rates ($\frac{1}{\tau_{\mathbf{q}\lambda}^{EP}}$). The PP scattering is generally considered as the main scattering mechanism of phonons in materials, and its rates can be obtained through the Fermi's golden rule[44, 45]:

$$\Gamma_{\lambda\lambda'\lambda''}^{\pm} = \frac{\hbar\pi}{4} \begin{Bmatrix} n'-n'' \\ n'+n''+1 \end{Bmatrix} \frac{\delta(\omega \pm \omega' - \omega'')}{\omega\omega'\omega''} |V_{\lambda\lambda'\lambda''}^{\pm}|^2, \quad (4)$$

The + (−) sign symbols the phonon absorption (emission) processes, and $V_{\lambda\lambda'\lambda''}^{\pm}$ is the phonon scattering matrix element. The EP scattering rates can be attained from the imaginary part of the phonon self-energy[46]:

$$\frac{1}{\tau_{\mathbf{q}\lambda}^{EP}} = \frac{2\pi}{\hbar} \sum_{mn,\mathbf{k}} |g_{mn}^{\lambda}(\mathbf{k},\mathbf{q})|^2 (f_{m\mathbf{k}+\mathbf{q}} - f_{n\mathbf{k}}) \delta(\varepsilon_{n\mathbf{k}} - \varepsilon_{m\mathbf{k}+\mathbf{q}} - \hbar\omega_{\mathbf{q}\lambda}), \quad (5)$$

$f_{n\mathbf{k}}$ is the Fermi-Dirac distribution function and $\varepsilon_{n\mathbf{k}}$ is the eigenvalue for electronic state ($n$, $\mathbf{k}$). The EP coupling matrix element $g_{mn}^{\lambda}(\mathbf{k},\mathbf{q})$ can be calculated through the density functional perturbation theory (DFPT) as[47]:

$$g_{mn}^{\lambda}(\mathbf{k},\mathbf{q}) = \sqrt{\frac{\hbar}{2M\omega_{\mathbf{q}\lambda}}} \langle \psi_{m\mathbf{k}+\mathbf{q}} | \partial V_{\mathbf{q}\lambda} | \varphi_{n\mathbf{k}} \rangle, \quad (6)$$

Here $\psi_{n\mathbf{k}}$ is the electronic wave function for initial state ($n$, $\mathbf{k}$) and $\psi_{m\mathbf{k}+\mathbf{q}}$ is that for the final state ($m$, $\mathbf{k}+\mathbf{q}$). $\partial_{\mathbf{q}\lambda} V$ is the first-order differential of Kohn-Sham potential due to atomic displacement associated with the phonon mode ($\mathbf{q}$, $\lambda$)[48].

**Computational details**

The accurate EP calculations within the framework of DFPT are performed by using the Quantum Espresso[49] package. The norm-conserving pseudopotentials[50] and Perdew−Burke−Ernzerhof (PBE) exchange-correlation functional within generalized gradient approximation[51] are employed. For the (non-) self-consistent calculation, the cutoff energy of plane wave is 100 Ry and the electronic energy convergence criterion is $10^{-10}$ Ry. The Broyden–Fretcher–Goldfarb–Shanno optimization method[52, 53] is used to optimize the crystal structure with the convergence criterion for the force acting on atoms less than $10^{-7}$ Ry/a.u. In order to calculate the EP scattering rates, the EP coupling matrix elements are first computed on the coarse **k(q)** grids by DFPT as embedding in Quantum Espresso, then interpolated to the fine ones by using the EPW[46] package. For $Ti_{0.25}Nb_{0.75}FeSb$ (12 atoms) and $Ti_{0.125}Nb_{0.875}FeSb$ (24 atoms) calculations, the coarse **k(q)** grids are 5×5×5 and 4×4×4, which are interpolated to 20×20×20 and 15×15×15 fine grids, respectively. And for the $Ti_{0.037}Nb_{0.963}FeSb$ (81 atoms) calculation, the simplified algorithm that proposed in this work is employed. To calculate the PP scattering rates, the Vienna *ab initio* Simulation Package (VASP)[54] and Phonopy code[55] are used to obtain the 2$^{nd}$-order IFCs. The 3$^{rd}$-order IFCs and $\kappa_L$s based on PP interactions are further calculated by VASP and ShengBTE[44] code, and the nearest neighbor atomic numbers are 5. Finally, we use the homemade code[56] to combine the PP and EP scattering rates, and calculate the lattice thermal conductivity according to Equation (2) and (3).

## Acknowledgements

This work is supported by the National Natural Science Foundation of China (Grant Nos. 52172216, 92163212, and 11975100) and Shanghai Technical Service Center of Science and Engineering Computing, Shanghai University. JY acknowledges the support from Hefei advanced computing center and Shanghai Engineering Research Center for Integrated Circuits and Advanced Display Materials.

## Author contributions

J. Xi and J. Yang designed the project. S. Dai and E. Dong performed the first-principles calculations. S. Zhang and Y. Sheng performed the physical and mathematical model. S. Dai and J. Yang analyzed the data and wrote the original manuscript, with input from S. Sun, L. Xi, and J. Snyder. supervised the project. All the authors reviewed and edited the manuscript.